\documentclass[12pt,preprint]{aastex}

\shorttitle{giant dark channel, filament channel, and chirality}
\shortauthors{Song et al.}

\begin{document}

\title{A giant dark channel across the solar equator consisting of two filament channels with different chiralities}

%A cusp structure at the junction of two filament channels with different chiralities
%A giant dark channel across the solar equator consisting of two filament channels with different chiralities

\author{Zhiping Song\altaffilmark{1}, Yijun Hou\altaffilmark{2,3}, and Jun Zhang\altaffilmark{2,3}}

\altaffiltext{1}{School of Physics and Materials Science, Anhui University, Hefei 230601, China}

\altaffiltext{2}{Key Laboratory of Solar Activity, National Astronomical Observatories,
Chinese Academy of Sciences, Beijing 100012, China; yijunhou@nao.cas.cn; zjun@nao.cas.cn}

\altaffiltext{3}{University of Chinese Academy of Sciences, Beijing 100049, China}

\begin{abstract}
Solar filaments are the largest magnetic structure that can be physically traced to the chromosphere.
The structure and evolution of solar filaments are important for our understanding of solar atmosphere
physics. In this work, we investigate a giant dark channel crossing the solar equator, which consists
of two filament channels with different chiralities. From 2016 April 22 to April 27, this giant dark
channel occupied the solar disk. Within this giant channel, a filament channel with dextral chirality
was detected in the northern hemisphere, and another filament channel with sinistral chirality was observed
in the southern hemisphere. At the junction of the two filament channels, a cusp structure was observed
associated with active region (AR) 12532 near the solar equator. The extrapolated three-dimensional magnetic
fields reveal that this cusp structure was composed of two sets of field lines belonging to two
different filament channels and was rooted on the AR positive-polarity fields. In addition, dark material
flows from filaments in the two channels to the cusp structure were detected as well as the flux emergence
and cancellation around the cusp footpoints. On 2016 May 21, after a solar rotation, the cusp structure had
disappeared, and the giant dark channel broke in the middle where another AR, 12546, had emerged
completely. We propose that the magnetic flux emergence and cancellation around the cusp region resulted
in the disappearance of the cusp structure and the break of the giant dark channel.
\end{abstract}


\section{Introduction}

Solar filaments consist of dense and cool plasma suspended in the hot solar corona with their lower part in the
chromosphere. On the solar disk, they show up in absorption as elongated and dark ribbons. But when seen above
the solar limb, they appear as bright features against the dark background and are called prominences. The terms
``filament'' and ``prominence'' are often interchangeably used in literature. In terms of the filaments' spatial
location on the Sun (Engvold 1998), we classified them as active region filaments (inside the active regions),
intermediate filament (at the border of active regions), and quiescent filament (on the quiet Sun). Filaments
are typically composed of a spine, barbs, and two extreme ends (Mackay et al. 2010). The earliest magnetograph
observations (Babcock \& Babcock 1955) showed that the filaments lie between different polarities of the Sun's
magnetic field, located above polarity inversion lines (PILs).

Filament channels are unique sites where filaments form in the sun, where the chromospheric fibrils are aligned
with the PIL (Martres et al. 1966; Gaizauskas 1998). These fibrils show the existence of the horizontal magnetic
field with the same direction on the two sides of the channel (Foukal 1971; Leroy et al. 1983; Martin et al. 1992).
A handedness property known as ``chirality'' has been discovered for filament channels, filaments, and their
overlying coronal arcades. Martin et al. (1994) proposed that the chirality of a filament channel is
defined as dextral if the axial magnetic field is directed to the right when viewed from the positive polarity
side of the channel. Conversely, it is defined as sinistral if the axial magnetic field viewed from the positive
polarity side is directed to the left. Since magnetic fields in filament channel are believed to extend to coronal
heights embedding the filaments, the chirality classification applies to the filaments as well. And there is a
one-to-one correlation that filaments in dextral channels have right-bearing barbs, and those in sinistral channels
have left-bearing barbs. Martin \& McAllister (1996) studied the coronal arcades overlying filaments and found that
left-skewed arcades always lie above dextral filaments while right-skewed arcades exist over sinistral filaments,
opposite to the right (left)-bearing structure of the filament barbs. Sheeley et al. (2013) proposed another
chirality identification method to infer the chirality of filament channels by using coronal cells in extreme
ultraviolet (EUV) 193 {\AA} images (Sheeley \& Warren 2012). Recently, Chen et al. (2014) and Ouyang et al. (2017)
determined the chiralities of erupting filaments according to the skewness of the conjugate filament drainage
sites. The filament chirality follows a hemispheric pattern that the dextral filaments dominate in the northern
hemisphere and sinistral in the southern hemisphere (Martin et al. 1994; Zirker et al. 1997; Pevtsov et al. 2003).

Previous works revealed that filaments or associated channels sometimes interact with each other and show
interesting characteristics in the chromosphere and low corona (Bone et al. 2009; Kumar et al. 2010; Li \&
Ding 2012; Jiang et al. 2014; Joshi et al. 2016; Xue et al. 2016; Zheng et al. 2017). Observations and
simulations have shown that these magnetic structures can reconnect and exchange their footpoint connectivity,
which is known as ``slingshot'' reconnection (Linton et al. 2001; Chandra et al. 2011; T{\"o}r{\"o}k et al.
2011; Jiang et al. 2013; Yang et al. 2017). Some authors have also presented observations where filaments of
the same chirality could merge to form one longer filament (Schmieder et al. 2004; Su et al. 2007; Joshi et al.
2014; Zhu et al. 2015). Numerical simulations also showed that two filaments occupying a single PIL in a
bipolar large-scale magnetic configuration could easily merged into a single prominence if their chiralities
were same and the axial magnetic fields were aligned (DeVore et al. 2005; Aulanier et al. 2006). Martin et al.
(1994) proposed that two filaments with different chiralities cannot merge into a single one and a cusp could
be formed at the ends of the two filaments where they meet. Schmieder et al. (2004) observed a cusp-shape
structure formed by the adjoining ends of two separate filaments with opposite chiralities.

From 2016 April 22 to April 27, the Solar Dynamics Observatory (\emph{SDO}; Pesnell et al. 2012) detected a
giant dark channel on the solar disk in EUV passbands, which spanned from the northern hemisphere to the
southern hemisphere. Within this giant dark channel, we identify a dextral filament channel in the northern
hemisphere and a sinistral filament channel in the southern hemisphere. Additionally, at the junction of
these two filament channels near the solar equator, a cusp structure was observed. Dark material
flows from the filaments within the two filament channels to the cusp structure were also detected. The
remainder of the paper is as follows: Section 2 deals with the description of the observations and
data analysis used in this study. The giant dark channel consisting of two filament channels with
different chiralities, the cusp structure at the junction of the two filament channels, the material flows
in the filaments within these two channels, and the evolutions of the cusp and the giant dark channel are
presented in Section 3. Finally, we conclude this work and discuss the results in Section 4.

\section{Observations and Data Analysis}

The Atmospheric Imaging Assembly (AIA; Lemen et al. 2012) on board the \emph{SDO} successively observes
multilayered solar atmosphere, including the photosphere, chromosphere, transition region, and corona,
in ten passbands, seven of which are in the EUV and observed with a cadence of 12 s and a spatial
resolution of 1.{\arcsec}2. From 2016 April 22 to April 27, a giant dark channel was observed over the
solar disk spanning from the northern hemisphere to the southern hemisphere in AIA EUV wavelengths.
During this period, we adopt the full-disk observations of the AIA 171 {\AA}, 193 {\AA}, and 211 {\AA}
to show this channel. Meanwhile, the line-of-sight (LOS) magnetograms from the Helioseismic and Magnetic
Imager (HMI; Schou et al. 2012) with a time cadence of 45 s and a pixel scale of 0.{\arcsec}5 are employed
to exhibit the underlying magnetic fields. The similar data are taken on 2016 May 21 to show this giant
channel after a solar rotation cycle. For the vector magnetic fields, the HMI data product called
Space-weather HMI Active Region Patches (SHARP; Bobra et al. 2014) are used as well. To investigate the
chromosphere configuration of the associated filaments, we also employ corresponding H$\alpha$ observations
from the Global Oscillation Network Group (GONG; Harvey et al. 2011). GONG has collected H$\alpha$ images
observed at seven sites around the world since mid-2010 and provides successive global H$\alpha$ observations
online.

In order to reconstruct the three-dimensional (3D) magnetic fields of the cusp structure, we perform
nonlinear force-free field (NLFFF) extrapolations based on the ``weighted optimization" method (Wiegelmann
2004; Wiegelmann et al. 2012). The vector magnetograms are preprocessed by a procedure developed by Wiegelmann
et al. (2006) to satisfy the force-free condition before being used as boundary condition. The NLFFF
extrapolation is performed in a box of 624 $\times$ 264 $\times$ 256 uniformly spaced grid points
(227 $\times$ 96 $\times$ 93 Mm$^{3}$).

\section{Results}
\subsection{A Giant Dark Channel Across the Solar Equator Consisting of Two Filament Channels with Different
Chiralities}

On 2016 April 24, a giant dark channel occupied the solar disk with a shape of fallen ``U'', which spanned
from the northern hemisphere to the southern hemisphere. Figure 1(a) shows the overview of this channel in
composite image of 171 {\AA}, 193 {\AA}, and 211 {\AA}. In these EUV passbands, the giant channel manifested
as an elongated and dark region. In panel (b), we exhibit the corresponding LOS magnetogram to check magnetic
fields underneath this channel. It's revealed that the elongated series of dark regions lay on the PIL
regions on a global scale. Specially, near the bottom of this ``U'' shape, that is the bend of this channel,
a distinct cusp-shaped structure was observed. Around this cusp structure, there was an active region
(AR) NOAA 12532. In the bend, the eastern side of the giant channel was dominated by the negative fields, and
the western side was dominantly positive, which was close to the positive polarity side of AR 12532.

A cusp structure, although not commonly seen, is widely considered to be formed between close ends of two filament
channels or filaments with opposite chiralities (Martin et al. 1994; Schmieder et al. 2004). In present work, we
detect a distinct cusp structure (see Figure 1(a)) in the bend of the dark channel near the solar equator on
days from 2016 April 22 to April 26, which implies there were two filament channels (see FC1 and FC2 in Figure 1)
with opposite chiralities within the elongated dark region and that they interacted with each other near
the solar equator. To look for evidence of filament channels within the giant dark regions, we apply the method
proposed by Sheeley et al. (2013) to infer the chirality of filament channels based on the coronal cells definition
in \emph{SDO}/AIA 193 {\AA} images. We select two areas located in the northern hemisphere and southern hemisphere
(see the white parallelograms in Figure 1(a)) to investigate the two filament channels, which are shown in Figures 2
and 3, respectively.

Figure 2 displays AIA 193 {\AA} image (top) and HMI LOS magnetogram (bottom) obtained from the selected region
outlined by the white dotted parallelogram in Figure 1(a) when it was around the central meridian on 2016 April 23.
The 193 {\AA} image used here has been constructed by normalizing the exposure time and averaging images over
a 5-minute interval to improve the signal-to-noise ratio. It's shown that the filament channel (FC1) is the
elongated dark region following approximately northeast-southwest along the PIL of the photospheric field.
Coronal cells are visible on the negative side of FC1 and affected by the spine magnetic field in the channel.
On the negative polarity side of this filament channel, these cells are stretched into tadpole shape with their
heads rooted in the majority polarity and their tails bending southwestward along the channel. We denote one of
these cells by blue circle (head) and arrow (tail). Because these cellular plumes are rooted in negative magnetic
field and their tails point to the southwest, we conclude that the axial magnetic field of FC1 shown in this
figure is directed to the northeast along the PIL. This corresponds to a dextral chirality, which is consistent
with the dominant chirality in the northern hemisphere.

Figure 3 shows a southern-hemisphere region marked by the white dashed parallelogram in Figure 1(a) on April 23.
In this region, the filament channel (FC2) is directed southeast-northwest along the PIL. Several coronal cells
are visible on the negative-polarity side of the channel. We mark one of these cells similar to that in Figure 2.
It is shown that this cell is rooted in concentration of negative flux and is stretched with its tail pointing
northwestward. This indicates a southeastward-directed axial field in FC2 and a sinistral chirality, as typical
for the southern-hemisphere filament channel.

\subsection{The Cusp Structure at the Junction of Two Filament Channels}

Using the coronal cells visible in AIA 193 {\AA} images, we find evidences of two filament channels within the
giant dark channel over the solar disk from April 22 to April 27. One is a long dextral FC1 in the northern
hemisphere. The other is a long sinistral FC2 in the southern hemisphere which crosses slightly the solar equator
into the northern hemisphere. At the junction of FC1 and FC2, where the southwest end of FC1 and the northwest
end of FC2 are close to each other, a cusp structure is observed. Figure 4 shows this cusp structure in 193 {\AA}
images (top panels), whose FOV is outlined by the green square in Figure 1. Superposing the AIA 193 {\AA} images
with the contours of the HMI LOS magnetograms, we notice that the southwest end of FC1 (see the dotted curve in
panel (a)) and the northwest end of FC2 (see the dashed curve) are all located on the positive fields of AR 12532.
Around 00:00 UT on April 23, two filaments were observed clearly within FC1 and FC2 through the H$\alpha$ image
in panel (d). During the following day, dark material flows from these two filaments to the cusp
structure were detected (see the black arrows in panel (e)). On April 24, the filaments in these two channels
as seen in the H$\alpha$ observation disappeared eventually (panel (f)).

Based on the photospheric vector magnetic fields at 10:00 UT on April 23, we extrapolate the 3D magnetic
topology of the cusp structure by using NLFFF modeling. Figures 5(a) and 5(b) show the extrapolation results from
the top view and side view, respectively. The cusp structure is depicted clearly, and consists of two sets of
field lines, which belong to two different filament channels (FC1 and FC2). It is shown that the cusp structure
is rooted on the positive-polarity fields of AR 12532. And the footpoints of the field lines belonging to FC1
(see the purple loop bundles) are on the south of that of the field lines belonging to FC2 (see the green loop
bundles).

To investigate the material flows from the filament in FC1 to the cusp structure, we extend the region
outlined by the green square in Figure 4(d) and show the evolution of this region in Figure 6 (also see the movie
1.mov attached to Figure 6). The filament in FC1 is distinct in 193 {\AA} images, with a barb extending
southeastward in the middle. From about 21:00 UT on April 22, the brightening appeared intermittently and repeatedly
near the foot of the barb (see the white arrow in panel (b)), where the consistent flux cancellation was detected as
well (see the blue circles in panels (d)-(f)). Meanwhile, the material of this filament moved successively toward the
cusp structure along the axis of FC1 (see the green arrows in panels (b) and (c)). The green domain ``A-B" in panel
(a) approximates the projected trajectory of the material flow in the plan of sky. We make a time-space
plot along this domain and display it in panel (g). It is shown that the filament material successively moved
away from the filament, forming repeated flows following separate trajectories. Then we delineate two trajectories
with green dotted curves and estimated their projected velocities. In these two trajectories, the flowing filament
material accelerated from $\sim$ 5.6 km s$^{-1}$ to $\sim$ 35.5 km s$^{-1}$ and from $\sim$ 5.2 km s$^{-1}$ to
$\sim$ 22.1 km s$^{-1}$, respectively.

In FC2, dark filament material flows toward the cusp structure was also detected (see the animation 2.mov
attached to Figure 7). Figure 7 displays the region outlined by the red square in Figure 4(d). The north part of FC2
and the dark flows (see green curved arrows in Figure 7(a2) and 7(b2)) were observed clearly in 193 {\AA} channel
from April 22 to 23. In addition, bright loops frequently appeared within FC2 and are denoted by the arrows
with different colors.

The results of NLFFF extrapolation reveal that the cusp structure was rooted on the positive-polarity fields
of AR 12532. To study the evolution of the magnetic fields around the footpoints of the cusp structure, we check HMI
LOS magnetograms from April 21 to 24 and show the results in Figure 8. Using the extrapolation as shown in
Figure 5, we approximate the location where the cusp structure was rooted and mark this region by red dashed
rectangles in Figures 8(a) and 8(d). On April 21, a negative-polarity magnetic patch (see the blue circles in
panels (a)-(c)) approached the positive fields where the cusp structure was rooted, and magnetic flux cancellation
kept occurring between them. On the other side of the cusp footpoints region, negative magnetic flux repeatedly
emerged and canceled with the surrounding fields with opposite polarity during April 23 (see the green rectangles
in panels (d)-(f)).

\subsection{The Giant Dark Channel After a Solar Rotation}

The giant dark channel occupied the solar disk from 2016 April 22 to April 27 until it rotated to the farside of the
Sun on April 28. After a solar rotation, this channel appeared again in the solar disk on 2016 May 21, and the shape
of the channel was a fallen ``U'' with a break in the position where the cusp was previously located but had
disappeared that time (see Figure 9). Figure 9(a) displays the solar disk in composite image of 171 {\AA}, 193 {\AA},
and 211 {\AA}. Two separate filament channels are distinguished unambiguously with one in the northern hemisphere and
the other in the southern hemisphere (see the black dashed curves). Combining the HMI LOS magnetogram (see panel (b)),
we notice that the filament channels was located above the PIL with a global scale. Around the bottom of this fallen
``U'', a strong active region NOAA 12546 completely emerged in the eastern side of the filament channel in
the southern hemisphere. Near the bend, the western side of the giant channel was dominated by the positive fields,
and the eastern side was dominantly negative, which was close to the positive polarity side of AR 12546.

To investigate the chiralities of the filament channels in different hemispheres, we choose two areas around these
two filament channels (see the squares in Figure 9(a)) and show them in Figure 10. Since the H$\alpha$ observations
exhibit clearly the filament barbs in these two regions, we identify the chiralities of these filament channels
according to the right-bearing or left-bearing filament barbs in the channels (Martin et al. 2008). Figures 10(a) and
10(b) exhibit the filament channel region in the northern hemisphere on May 21. In 171 {\AA} image of panel (a), we
contour the negative and positive magnetic fields with blue and red curves while the green curves are the contour of
the filament fragments as seen in the H$\alpha$ observation (see panel (b)). The filament here was located in the
west-east area above the PIL. In panel (b), the dextral filament barbs in this region are denoted by the white
arrows in H$\alpha$ image. Another filament channel region in the southern hemisphere is shown in Figures 10(c) and
10(d). Panel (d) reveals that the filament barbs in this channel region are sinistral. Thus the two filament
channels exhibit the same chiralities as in the previous rotation.

\section{Summary and Discussion}

Employing the successive observations from the \emph{SDO} and GONG, we investigate a giant dark channel occupying
the solar disk from 2016 April 22 to April 27. Within this dark channel, we detect two filament channels. One is a
long dextral filament channel (FC1) in the northern hemisphere, and the other one is a long sinistral filament
channel (FC2) in the southern hemisphere which overlaps slightly into the northern hemisphere. These two
filament channels with different chiralities come together near the solar equator, forming a cusp structure associated
with AR 12532. The extrapolated 3D magnetic fields confirm the existence of this cusp structure and reveal that it
was rooted on the positive-polarity fields of AR 12532. Moreover, dark material flows from the filaments within
these two filament channels to this cusp were detected. Intermittent magnetic flux emergence and cancellation were
also observed around the footpoints of the cusp structure. On 2016 May 21, after a solar rotation, the cusp structure
had disappeared, and the giant dark channel broke in the middle where another AR, NOAA 12546, had emerged
completely. The chiralities of the two filament channels within this dark channel remained as dextral in the
northern hemisphere and sinistral in the southern hemisphere, as in the previous solar rotation.

From 2016 April 22 to April 27, the giant dark channel extended from the northern hemisphere to the southern
one, occupying the solar disk. Existence of such a large-scale region in the solar disk suggests the global order of
the solar magnetic field. This elongated series of dark regions lie on the PIL with global scale, within which two
filament channels exist: dextral FC1 in the northern hemisphere and sinistral FC2 in the southern hemisphere. The
northwestern side of FC1 was dominantly positive magnetic flux and the southeastern side negative magnetic flux.
Meanwhile, the dominant flux in the southwestern side of FC2 was positive and the northeastern side negative.
From 2016 April 22 to April 27, AR 12532 was located on the western side of FC1 near its southwest end, where the
channel was reinforced by the positive flux elements of this AR. A similar phenomenon was reported by
Anderson \& Martin (2005) as well. They suggested that the flux of new active regions supplements the dominant
polarity of the channel on each side, sustaining the channel for a long time. In present work, the giant dark
channel existed for a long time and appeared again in 2016 May. After a solar rotation, the giant channel broke
in the middle, where another active region NOAA 12546 emerged completely on the east side of FC2 near its
northwest end. The positive flux of this AR canceled the flux sustaining the channel. According to the Hale's law
of AR magnetic orientation, for the 24 solar cycle, the following polarity of an AR in the northern
hemisphere would be positive and the leading polarity of an AR in the southern hemisphere would be positive. As a
result, unless an AR exactly forms on the axis of FC1 in the northern hemisphere, the AR positive polarity
will be closer to FC1 on its western side and the AR negative polarity will be closer on the eastern side, which
would statistically sustain FC1. However, for the FC2 in the southern hemisphere, the surrounding ARs would
statistically cancel the flux sustaining the channel. We propose that the continuous sustainment of flux
elements from the nearby ARs in the northern hemisphere (such as AR 12532) may support the continued existence
of FC1 from 2016 April to 2016 May. However, if some ARs emerge beside this channel in the southern hemisphere
(such as AR 12546), the flux from these neighboring ARs would cancel the flux sustaining FC2, the northern part
of which had been destroyed by May 21. This process may eventually lead to the disappearance of FC2.

Previous studies reveal a statistical preference for dextral chirality in the northern hemisphere and sinistral chirality
in the southern hemisphere (Rust 1967; Martin et al. 1994; Bernasconi et al. 2005). This hemispheric pattern is unusual,
as it is exactly opposite to the result of differential rotation acting on a north-south coronal arcade overlying an
east-west PIL (Mackay \& van Ballegooijen 2005). To explain this phenomenon, two groups of theories were developed,
i.e., the surface effects (Kuperus 1996; Zirker et al. 1997; Martens \& Zwaan 2001) and the subsurface effects (Rust
\& Kumar 1994; Priest et al. 1996). Zirker et al. (1997) proposed that the combination of supergranular motions,
differential rotation, and magnetic reconnection may lead to the hemispheric pattern. Lim \& Chae (2009) suggested
that the chirality of intermediate filaments may result from the magnetic helicity of their associated active regions.
In present work, the chirality distribution of the filament channels was consistent with the hemispheric pattern
during the solar rotation from 2016 April to May. Our observations of the giant dark channel and the chiralities of
the two filament channels within this giant channel suggest the existence of global order of the solar magnetic field.

At the junction of the two filament channels (FC1 and FC2) with different chiralities, we observe a cusp structure.
Cusp structures between close ends of two filaments or filament channels have rarely been reported (Martin et al. 1994;
Schmieder et al. 2004). In the present work, for the first time, we investigate the magnetic topology, associated dynamic
activities, and the evolution of underlying magnetic fields of the cusp structure. The NLFFF extrapolation reveals
that the cusp structure reported in our work consists of two sets of field lines and is rooted in the positive-polarity
fields of AR 12532. These two sets of field lines respectively belong to FC1 and FC2. In addition, the footpoints of
the field lines belonging to northern FC1 are on the south of that of the field lines belonging to southern FC2 (see
Figure 5). The reverse locations of the footpoints of the two sets of field lines imply that rotational motion of the
AR magnetic patch where the footpoints are rooted could have contributed to the formation of this cusp structure
(Kumar et al. 2010).

Dynamics of the solar filaments have been studied for several decades (Schmieder et al. 1985, 2017).
Here we also detect dark material flows from the filaments in FC1 and FC2 to the cusp. In FC1, the intermittent
brightening in 193 {\AA} wavelength and magnetic flux cancellation occurred near the foot of one filament barb (see
Figure 6). This implies that the flux cancellation destabilized the filament in FC1 and disturbed
the filament material (Deng et al. 2002; Li et al. 2015; Schmieder et al. 2006), which then moved successively toward
the cusp structure along the axis of FC1. Similarly, the dark flows were observed in FC2 as well as bright loops.

The successive \emph{SDO} observations show that the cusp structure disappeared on 2016 May 21 after a solar
rotation. Checking the evolution of the magnetic fields around the footpoints of the cusp structure from from April
21 to 24, we notice that magnetic flux emergence and cancellation kept occurring around the cusp footpoints region.
Moreover, when the cusp structure disappeared after a solar rotation, AR 12546 had emerged completely around the
north part of FC2, where the giant dark channel broke. Therefore, we suggest that the intermittent magnetic flux
emergence and cancellation in the cusp region destroyed the magnetic structure of the cusp structure and eventually
resulted in the disappearance of the cusp structure and the break of the giant dark channel.

\acknowledgments{
The authors are grateful to the anonymous referee for valuable suggestions.
The data are used courtesy of the \emph{SDO} and GONG science teams. \emph{SDO} is a mission of NASA's Living With
a Star Program. This work is supported by the National Natural Science Foundations of China (U1531113, 11533008,
11790304, 11773039, 11673035, 11673034, 11873059, and 11790300) and Key Programs of the Chinese Academy of Sciences
(QYZDJ-SSW-SLH050).
}

{}
\clearpage

\begin{figure}
\centering
\includegraphics [width=.98\textwidth]{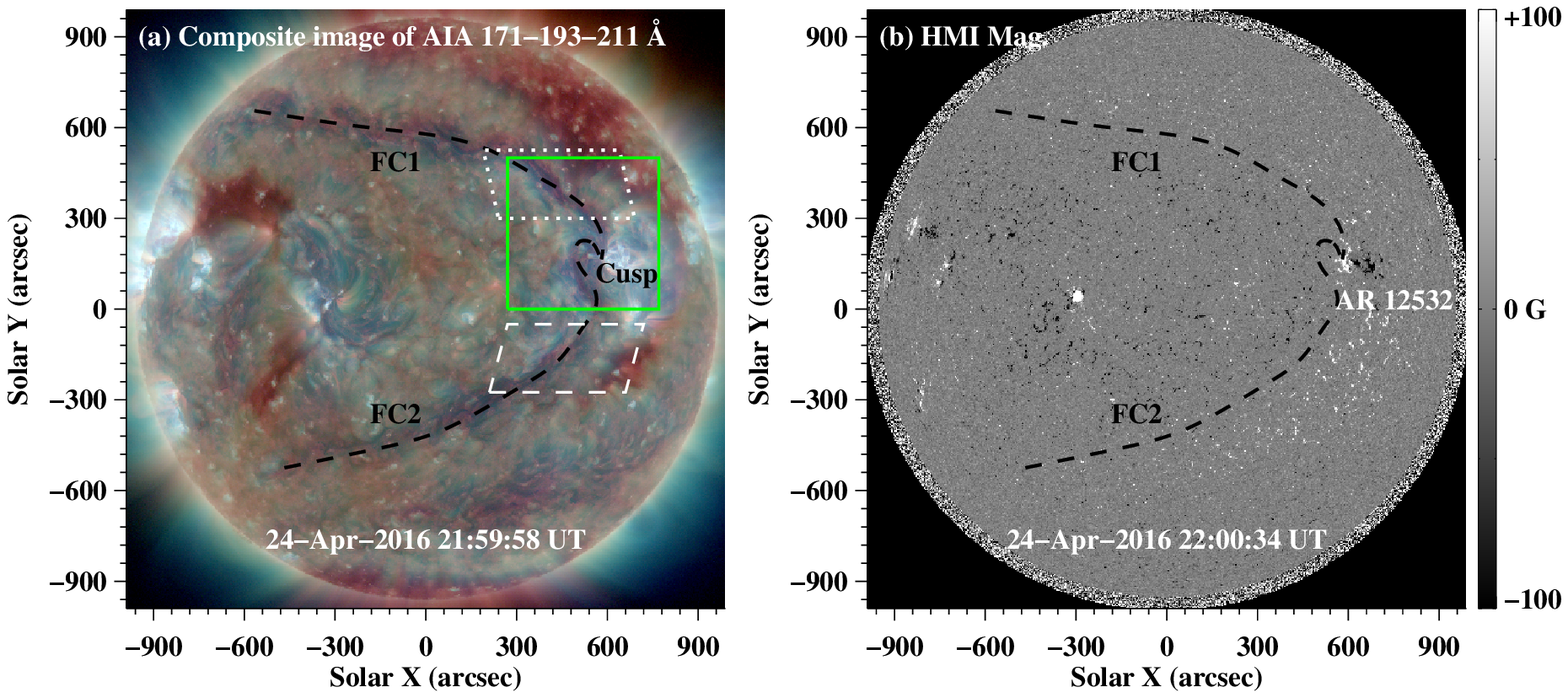}
\caption{
Composite image of \emph{SDO}/AIA 171 {\AA}, 193 {\AA}, and 211 {\AA} and \emph{SDO}/HMI LOS magnetogram showing
the giant dark channel over the solar disk on 2016 April 24 in EUV passbands and the underneath magnetic fields.
The black dashed curves in panel (a) represent the axis of the filament channels and are duplicated in panel (b).
The white dotted, dashed parallelograms, and green square in panel (a) approximate the field of views (FOVs)
of Figures 2, 3, and 4.
}
\label{fig1}
\end{figure}

\begin{figure}
\centering
\includegraphics [width=0.8\textwidth]{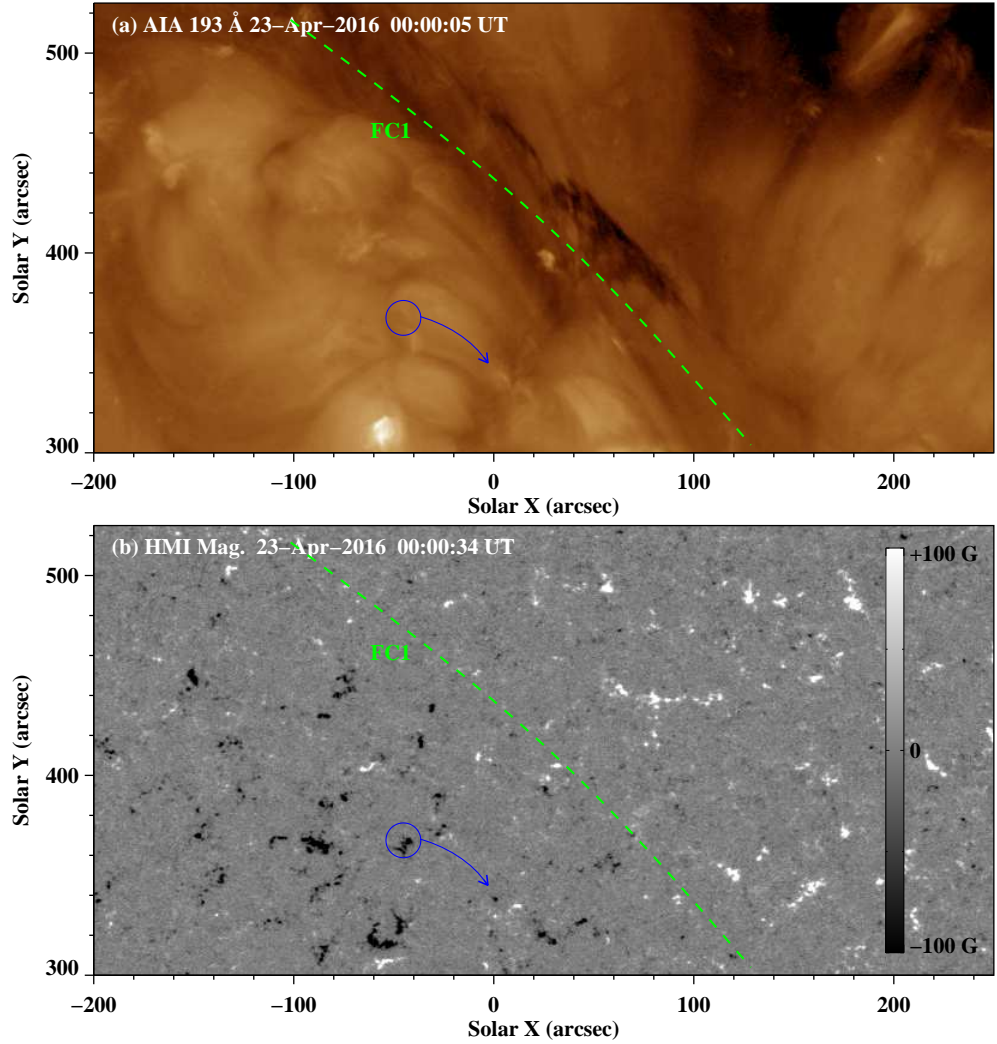}
\caption{
AIA 193 {\AA} image and HMI LOS magnetogram showing the filament channel in the northern hemisphere
within the giant dark channel on 2016 April 23. On the southeast side of this channel, coronal cells are
distorted into tadpole shapes with their tails pointing to the southwest along the channel. The blue circle
marks the head of one of these coronal cells, which is rooted in the majority polarity. And the blue arrow
points out the tail direction of this cell. This corresponds to a northeastward-directed field along the filament
channel and a dextral chirality.
}
\label{fig2}
\end{figure}

\begin{figure}
\centering
\includegraphics [width=0.8\textwidth]{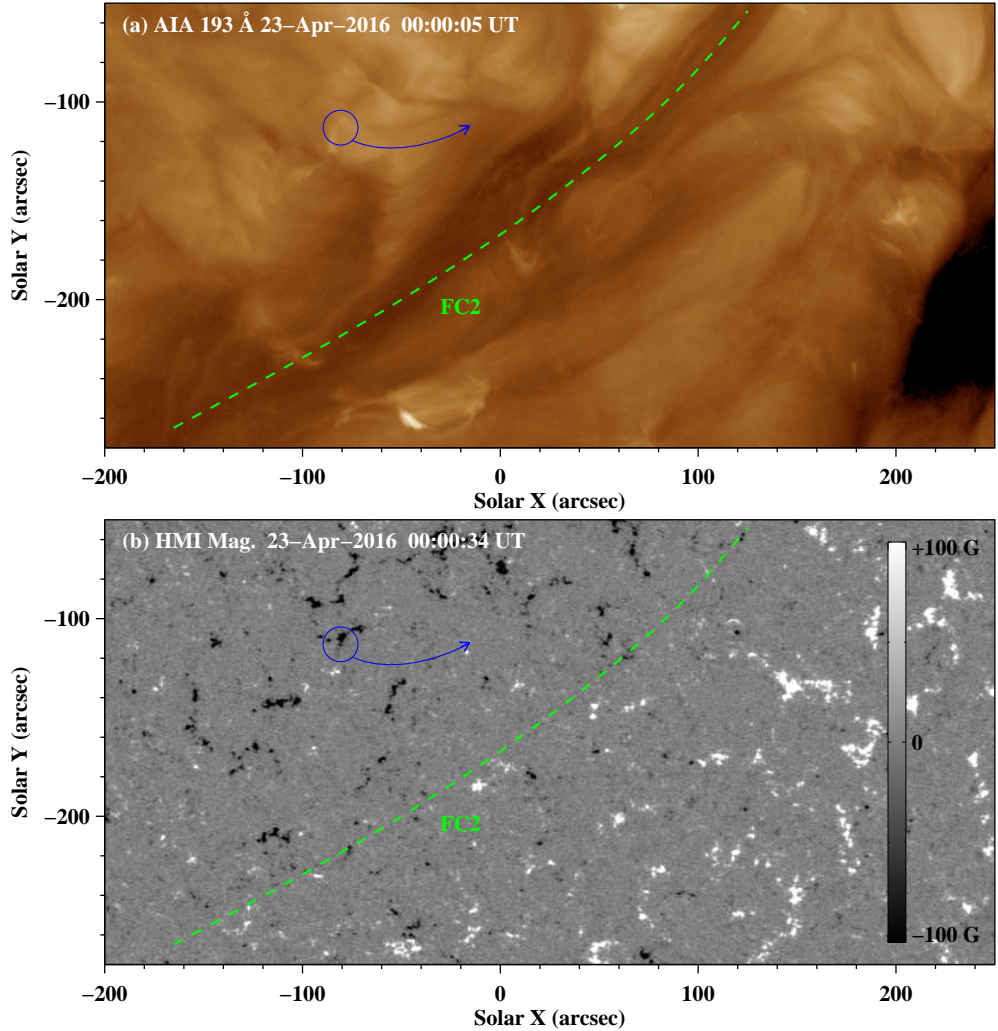}
\caption{
Similar to Figure 2, but for the filament channel in southern hemisphere within the giant dark channel.
Coronal cells on the northeast side of the filament channel are still distorted into tadpoles
with their heads rooted in the majority polarity and their tails oriented in northwest along the channel. One
of these cells is marked by the blue circle and arrow. This corresponds to a southeastward-directed field
along the filament channel and a sinistral chirality.
}
\label{fig3}
\end{figure}

\begin{figure}
\centering
\includegraphics [width=.98\textwidth]{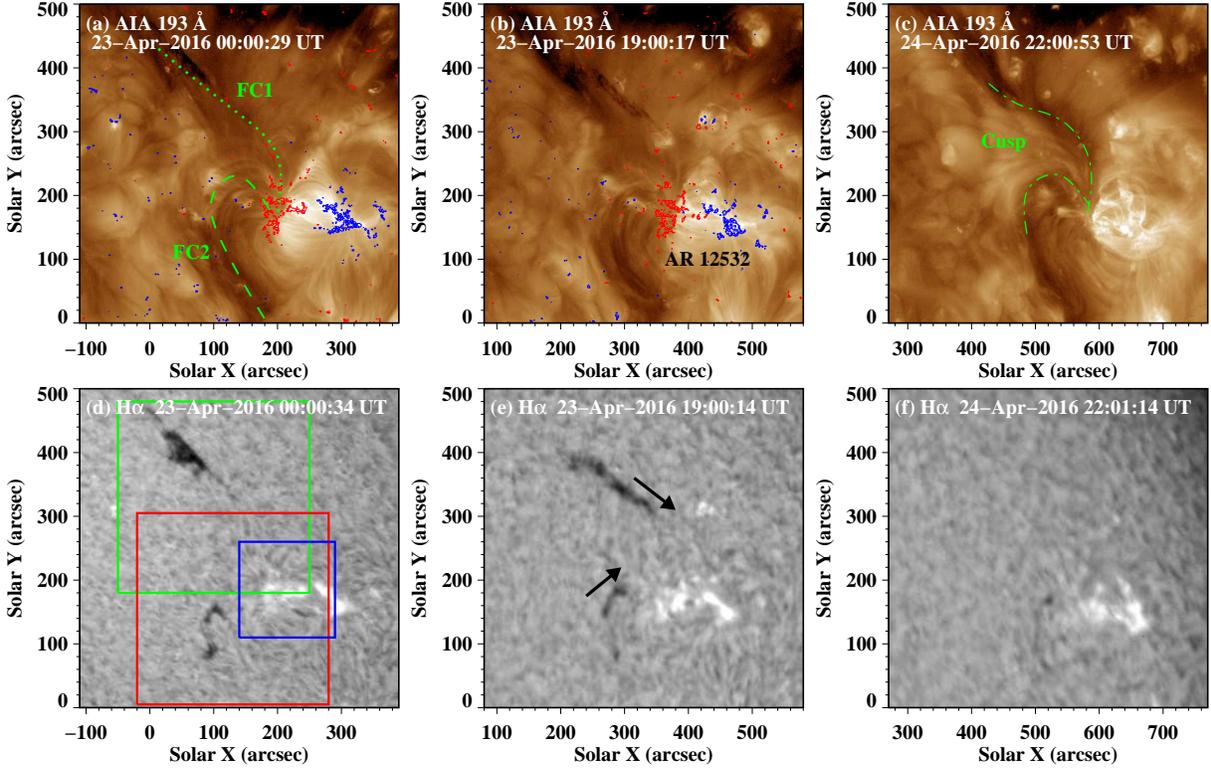}
\caption{
Observations of 193 {\AA} and H$\alpha$ wavelengths showing the cusp structure at the junction of the
two filament channels (FC1 and FC2) and two filaments within FC1 and FC2.
Panels (a)-(c): sequence of AIA 193 {\AA} images. The red and blue curves are contours of the LOS magnetograms
at +200 and -200 G, respectively. The axes of FC1 and FC2 are delineated by the green dotted and dashed curves
in panel (a).
Panels (d)-(f): sequence of H$\alpha$ images. The H$\alpha$ image in panel (d) is obtained from the Learmonth
Solar Observatory, and the images in panels (e) and (f) are taken by the Mauna Loa Solar Observatory.
The green, red, and blue squares in panel (d) outline the FOVs of Figures 6, 7, and 8, respectively.
The black arrows in panel (e) denote the directions of filament material flows within FC1 and FC2.
}
\label{fig4}
\end{figure}

\begin{figure}
\centering
\includegraphics [width=.6\textwidth]{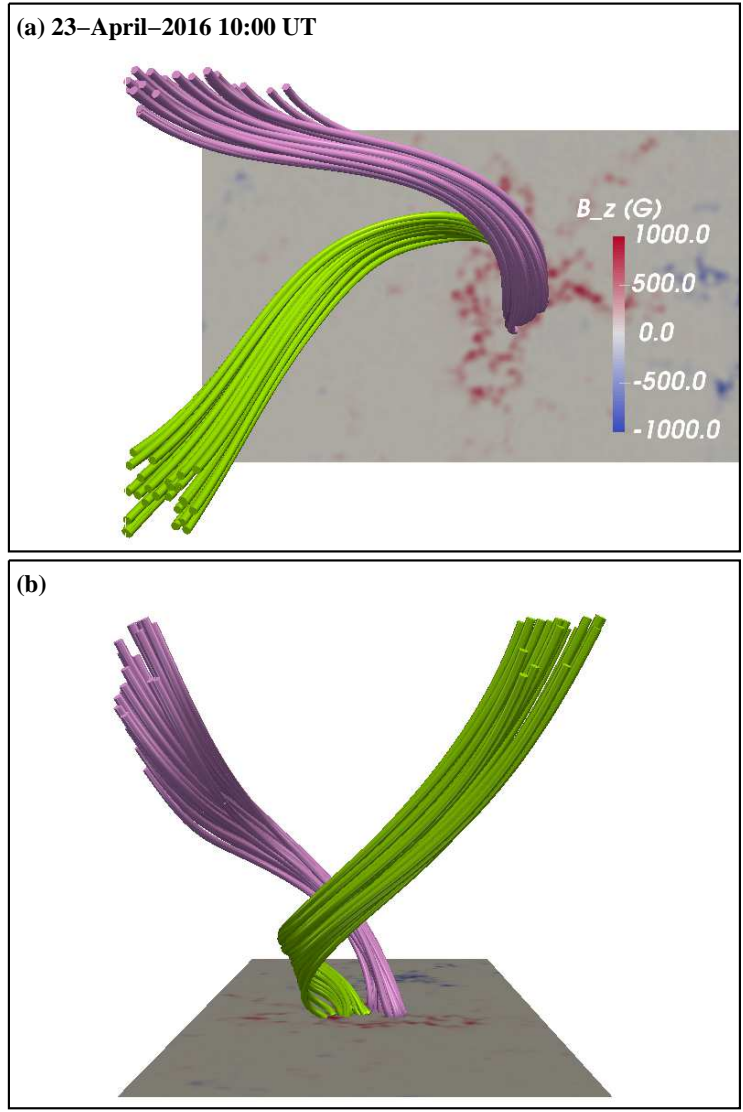}
\caption{
Cusp structure revealed by NLFFF modeling at 10:00 UT on April 23. Panels (a) and (b) show the top view
and side view of this cusp structure, respectively. The photospheric vertical magnetic field (B$_z$) is shown as
background with a saturation of $\pm$1000 G. The purple and green loop bundles represent two sets of
field lines belonging to FC1 and FC2, respectively.
}
\label{fig5}
\end{figure}

\begin{figure}
\centering
\includegraphics [width=.86\textwidth]{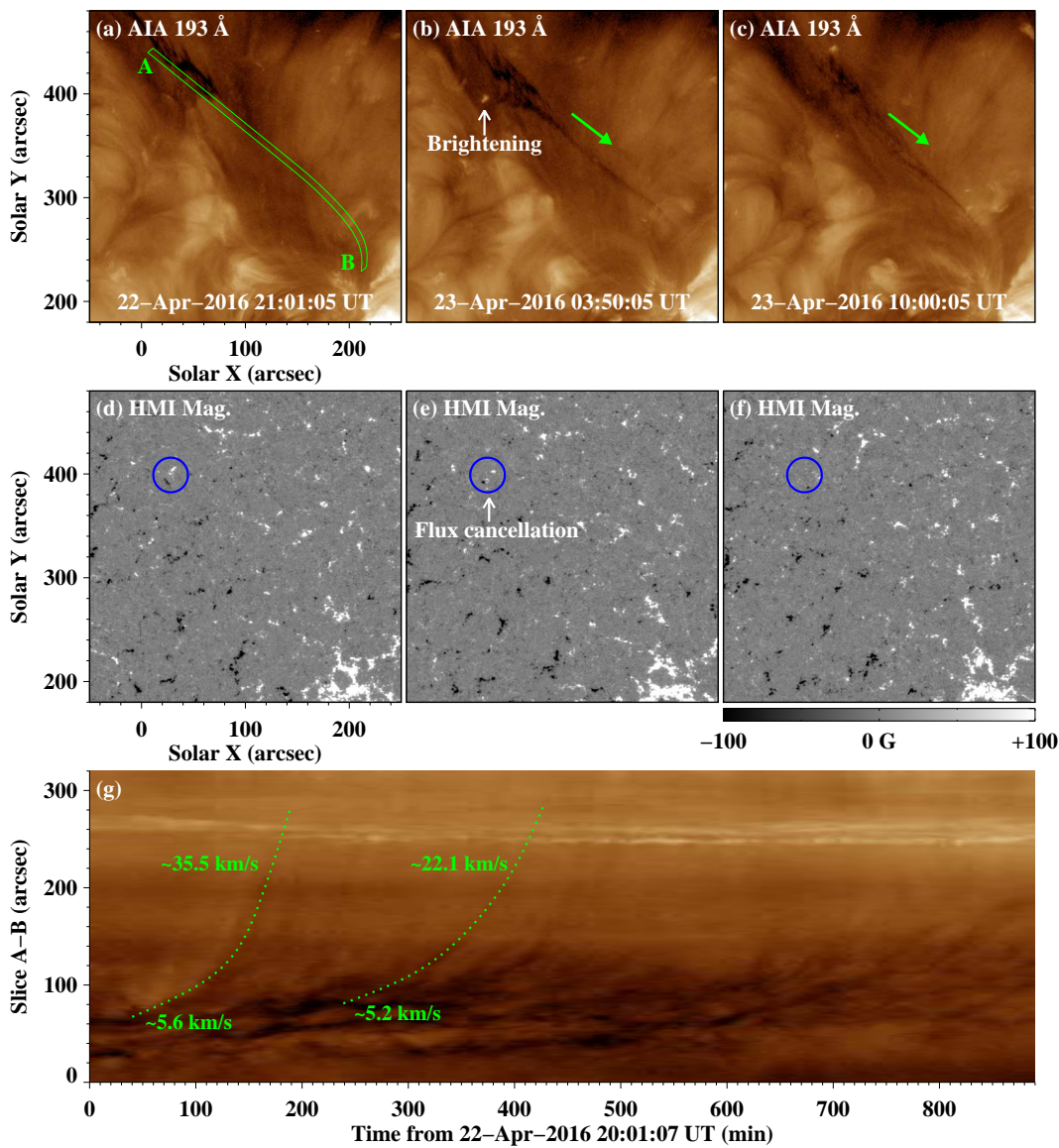}
\caption{
The filament material flows from the filament in FC1 to the cusp structure.
Panels (a)-(c): a series of 193 {\AA} images displaying the material flows. The green domain ``A-B'' in
panel (a) is along the projected trajectory of the flows, and the green arrows in panels (b) and (c) point
out its direction. The emission strengthening around the foot of the filament barb is marked by the white arrow
in panel (b).
Panels (d)-(f): sequence of HMI LOS magnetograms showing the evolution of the underlying magnetic fields. The blue
circles label the position corresponding to the brightening in 193 {\AA} images, where magnetic flux cancellation
is detected.
Panel (g): time-space plot along domain ``A-B'' in 193 {\AA} channel. The green dotted lines approximate the
projected trajectories of the flowing filament material.
(An animation (1.mov) of the 193 {\AA} images and HMI LOS magnetograms shown in this figure is available in
the online edition.)
}
\label{fig6}
\end{figure}

\begin{figure}
\centering
\includegraphics [width=.86\textwidth]{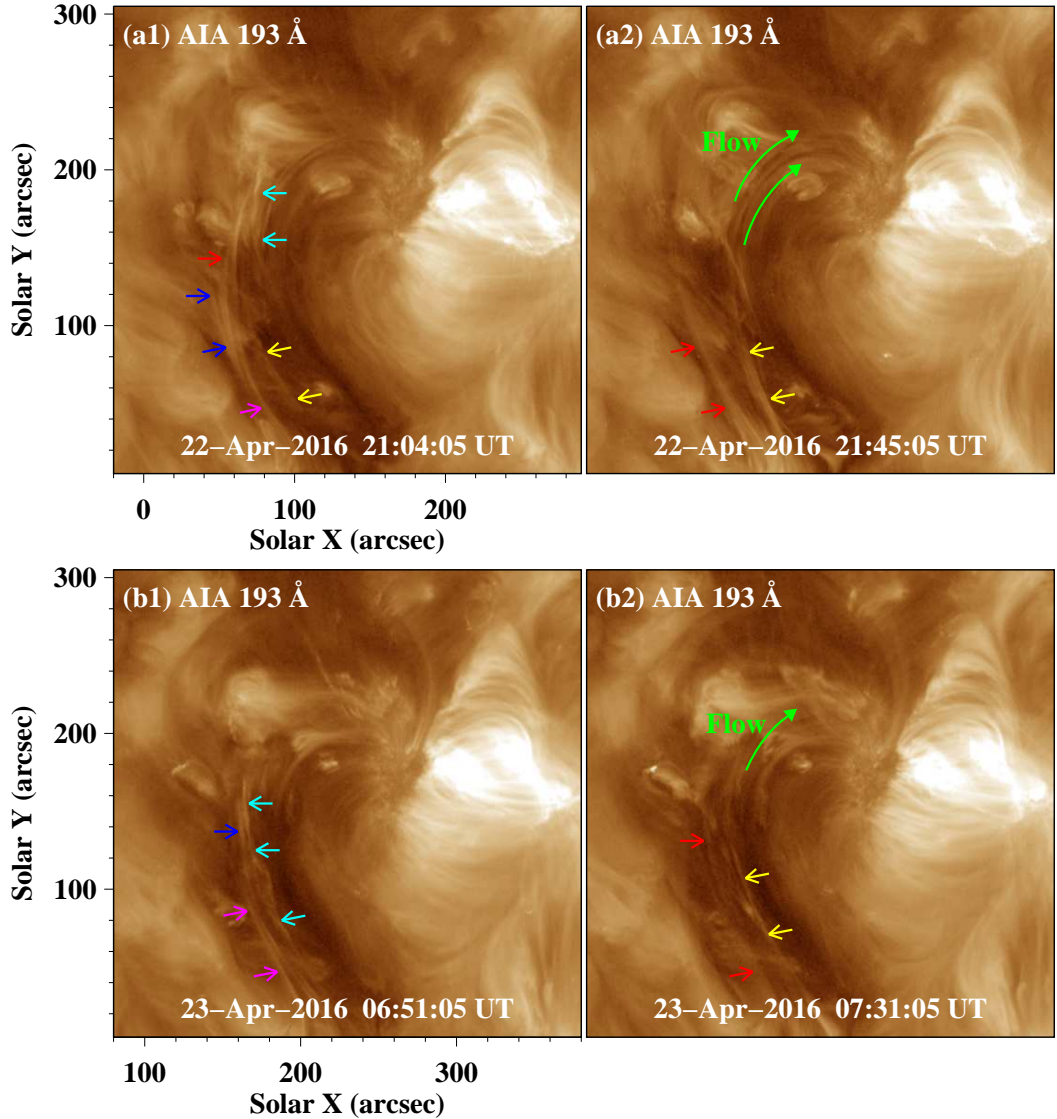}
\caption{
AIA 193 {\AA} images displaying the bright loops and dark filament material flows within FC2. Panels (a1) and
(a2) clearly show the loops and flows around 21:30 UT on April 22. The arrows with various colors denote different
bright loops. The green curved arrows in panel (a2) mark the dark filament flows. Panels (b1) and (b2) are similar to
panels (a1) and (a2), but for the time around 07:10 UT on April 23.
(An animation (2.mov) according to panels (a1) and (a2) is available in the on-line journal.)
}
\label{fig7}
\end{figure}

\begin{figure}
\centering
\includegraphics [width=.9\textwidth]{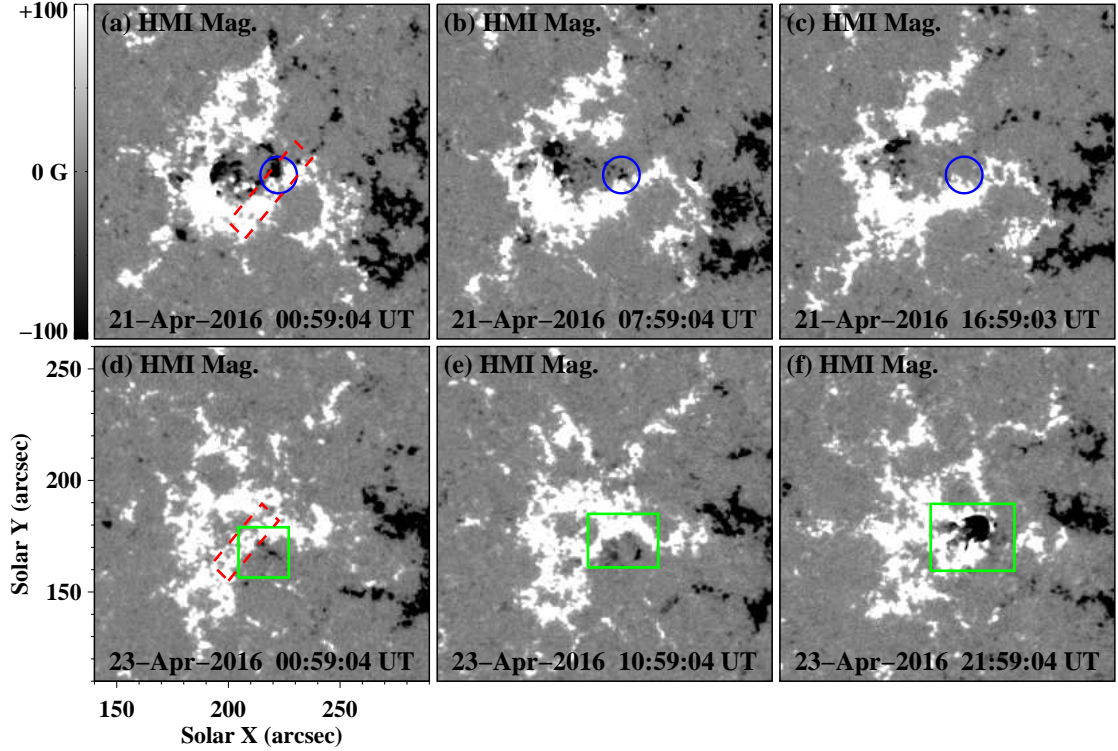}
\caption{
Time sequence of HMI LOS magnetograms showing the evolution of the magnetic fields around the footpoints of
the cusp structure. The red dashed rectangles in panels (a) and (d) approximate the location where the cusp structure
is rooted according to the result of NLFFF modeling shown in Figure 5. Panels (a)-(c) and (d)-(f) exhibit two processes
of magnetic flux emergence and cancellation on April 21 and 23, respectively. The blue circles and green rectangles
outline regions where the flux emergence and cancellation occurred.
}
\label{fig8}
\end{figure}

\begin{figure}
\centering
\includegraphics [width=.98\textwidth]{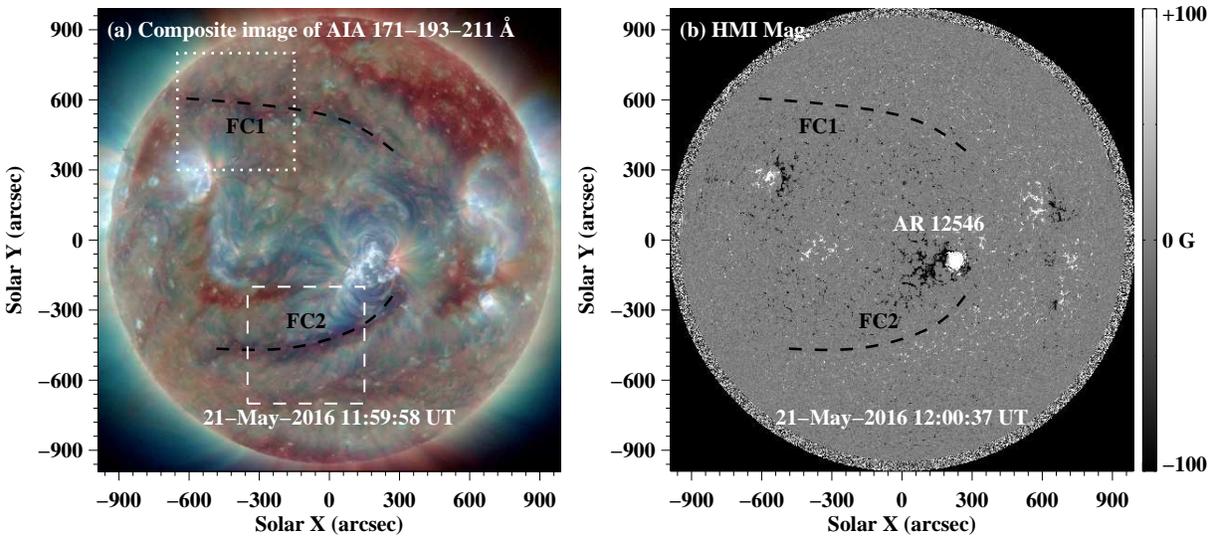}
\caption{
Composite image of AIA 171 {\AA}, 193 {\AA}, and 211 {\AA} and HMI LOS magnetogram showing the giant dark channel and
the corresponding magnetic fields, after a solar rotation on 2016 May 21. Two filament channels are delineated by the
black dashed curves in panel (a), which are duplicated in panel (b). The white dotted square in panel (a) outlines the
FOV of Figures 10(a)-10(b) and the dashed one represents the FOV of Figures 10(c)-10(d).
}
\label{fig9}
\end{figure}

\begin{figure}
\centering
\includegraphics [width=0.86\textwidth]{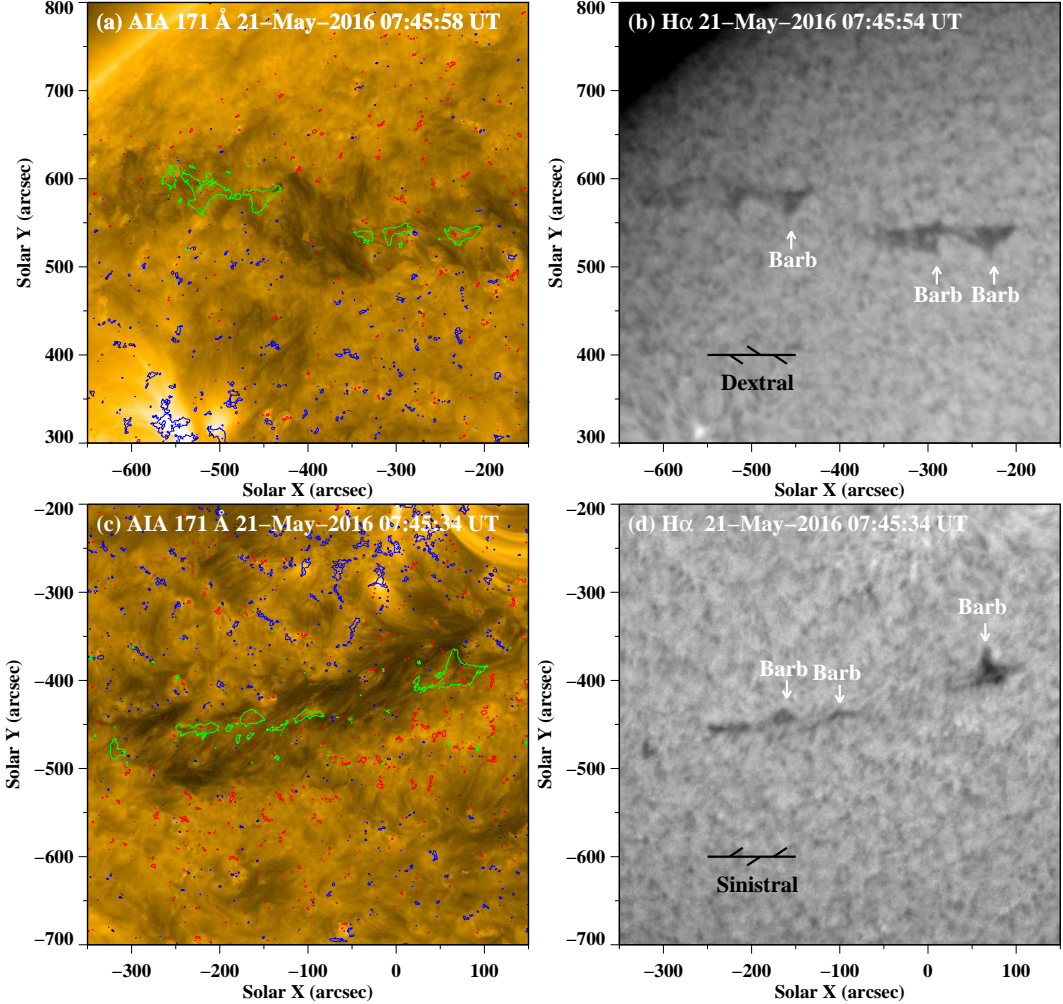}
\caption{
Two filament channels in the northern and southern hemispheres, respectively.
Panels (a)-(b): AIA 171 {\AA} image and H$\alpha$ image from the Udaipur Solar Observatory showing the filament channel
in the northern hemisphere. In panel (a), the blue and red curves show the negative and positive magnetic fields beside
the channel, respectively. And the green curves outline the filament fragments inside the channel. The dextral barbs
of the filament are marked in panel (b).
Panels (c)-(d): AIA 171 {\AA} image and H$\alpha$ image from the Learmonth Solar Observatory displaying the filament
channel in the southern hemisphere. The features marked here are similar to those in panels (a)-(b). However, the
chirality of the filament in this filament channel is sinistral.
}
\label{fig10}
\end{figure}

\end{document}